\documentclass[submission,copyright,creativecommons]{eptcs}
\providecommand{\event}{TARK XVIII 2021} 
\usepackage{breakurl}             
\usepackage{underscore}           
\usepackage{amsthm, amsmath, amsfonts, mathtools, amssymb,stmaryrd, proof} 
\usepackage{graphicx}
\usepackage{sgame, tikz-qtree}
\usetikzlibrary{trees, calc} 
\usepackage{multirow}
\theoremstyle{definition}

\title{Epistemic Modality and Coordination under Uncertainty\thanks{This work has received funding from the European Research Council (ERC) under the European Union’s
Horizon 2020 research and innovation program (Grant Agreement No 758540) within the project \textit{EXPRESS: From the Expression of Disagreement to New Foundations for Expressivist Semantics}. Special thanks to Luca Incurvati, Le\"ila Bussi\`ere, and to the participants of the 2020 Dutch Research School in Philosophy conference for discussion.}}
\author{Giorgio Sbardolini
\institute{ILLC, University of Amsterdam\\
Amsterdam, The Netherlands}
\email{g.sbardolini@uva.nl}
}
\def\titlerunning{Coordination Under Uncertainty}
\def\authorrunning{G. Sbardolini}
\begin{document}
\maketitle

\begin{abstract}
Communication facilitates coordination, but coordination might fail if there's too much uncertainty. I discuss a scenario in which vagueness-driven uncertainty undermines the possibility of publicly sharing a belief. I then show that asserting an epistemic modal sentence, `Might $\phi$', can reveal the speaker's uncertainty, and that this may improve the chances of coordination despite the lack of a common epistemic ground. This provides a game-theoretic rationale for epistemic modality. The account draws on a standard relational semantics for epistemic modality, Stalnaker's theory of assertion as informative update, and a Bayesian framework for reasoning under uncertainty.
\end{abstract}

\noindent Shiv and Logan want to spend time together over the coming weekend. They prefer to go to the beach if it will be sunny and to a caf\'e if it will be raining, but they will only go to either place if the other goes. Their predicament is the familiar one of a coordination problem \cite{lew69}. In a variant known as the signalling game, Shiv and Logan coordinate by sending a signal, i.e.\ an utterance that reveals information that is initially available to only one of the players \cite{rabin90,RabinFarrell1996,stalnaker2006-cheaptalk,sky10}.

It's relatively well-understood how Shiv and Logan coordinate if the relevant information (it will be sunny, it will be raining) is public. Roughly, $q$ is public information within a group just in case all members of the group believe that $q$, all believe that all believe that $q$, all believe that all believe that all believe that $q$, and so on. Sometimes, however, a belief may fail to be public. For example: Shiv thinks that it will be raining, but she's not very confident, and indeed she expects that, reasonably, Logan thinks that it will be sunny. In this case Shiv does not believe that she and Logan share the belief that it will be raining. The belief fails to be public. Is there still a way for Shiv and Logan to coordinate for the weekend, despite the uncertainty?

In this paper, I provide a rational reconstruction of the use of epistemic modals in a signalling game. I will present a game-theoretic rationale for epistemic possibility talk: revealing one's uncertainty to the interlocutors can improve one's expected utility despite lack of public information. I will employ a general Bayesian framework for reasoning under uncertainty \cite{franke11,fjr12,Goodman2016-GOOPLI-2,Lassiter2017-LASAVI}. The model will show that, in conditions of uncertainty specified below, rational agents can improve their chances of coordination by means of sentences that make an epistemic hedge. 

My focus is on sentences of the form `Might $\phi$', where an epistemic possibility operator takes wide scope. Epistemic possibility modal auxiliaries and adverbs, and expressions of close kin (\textit{might, perhaps, maybe, possibly, probably}) are natural resources to employ in case coordination is challenged by the lack of an epistemic common ground. If Shiv uttered (\ref{epist1}) in the story above, for example, she would be naturally understood as suggesting to go to a caf\'e for the weekend---almost as if she simply asserted that it will be raining, while at the same time hedging that assertion.
\begin{enumerate}
     \item\label{epist1} It might be raining.
    \begin{enumerate}
        \item\label{e1a} Let's stay in.
        \item\label{e1b} Let's go out.
    \end{enumerate}
\end{enumerate}
In her context, it is very natural for Shiv to continue (\ref{epist1}) as in (\ref{e1a}), much less natural (and not even so coherent) to continue as in (\ref{e1b}). Therefore, even if the speaker is not in a position to outright assert that it will rain, the interlocutors might still coordinate on going to a caf\'e by means of something less committal than that assertion.

\section{Failures of Coordination}\label{failure}
A game is a set of players $I$, and sets of actions $x_i$ and utility functions $u_i$ for each player $i$ in $I$. In a coordination game between two players, each having two actions, $a$ and $b$, the players' utility functions are summarized in Table \ref{coord}. Shiv (the Column player $S$) prefers to stay in, $a$, if Logan (the Row player $L$) stays in, and prefers to go out, $b$, if Logan goes out. Same for Logan.

\begin{table}[htb!]
    \centering
    \begin{tabular}{l c c c}
 &   & \multicolumn{2}{c}{$S$} \\ \cline{2-4}
 & \multicolumn{1}{|l}{}   & $a$ & $b$ \\
 \multirow{2}{*}{$L$} & \multicolumn{1}{|l}{$a$}    & $1,1$ & $0,0$ \\
   & \multicolumn{1}{|l}{$b$}     & $0,0$ & $1,1$ 
    \end{tabular}
    \caption{Coordination Game}
    \label{coord}
\end{table}

This particular assignment of preferences makes the agents \textit{indifferent} as to whether they stay in or go out so long as each does what the other does. How do they coordinate? Either could commit to an action, and then inform the other about their commitment. But if they are indeed indifferent, they might as well toss a (fair) coin, and resolve to stay in if and only if it lands heads. Suppose furthermore that only one of them, $S$, can see the coin: she will then report to $L$ the outcome of the coin toss by sending a signal (`It's tails!'). Coordination is then easily achieved \cite{rabin90,stalnaker2006-cheaptalk}.

Shiv and Logan can coordinate by ``pivoting'' on the weekend being rainy, rather than the coin landing tails, as well as on any other proposition. Let $q$ and $\bar{q}$ be two mutually exclusive propositions. The agents mutually know that they prefer $a$ if $q$ and $b$ if $\bar{q}$. If Shiv believes that $q$, she may signal so, and thereby share her belief with Logan. Neither Shiv nor Logan has reason to deceive the other, since either receives a positive payoff just in case the other does (as shown in Table \ref{coord}). Therefore, once a belief is shared by signalling, it typically becomes public: both believe it, both believe that both believe it, and so on. In a coordination game, if $q$ is public between Shiv and Logan, the rational (i.e.\ utility maximizing) choice for both is $a$.\footnote{This is version of the traditional signalling game described by David Lewis \cite{lew69}, with a couple of qualifications. (i) Lewis talked about common knowledge, rather than common belief, but the stronger condition doesn't add much at this stage: people can coordinate on something false, so long as enough people believe it. (ii) Lewis worked with the notion of a Nash equilibrium, but the idea of solving the game by reference to an event with an independent prior probability (the coin lands heads, the weekend will be rainy) leads in fact to a generalization known as correlated equilibrium, with which I work in the current paper \cite{aumann1987,Vanderschraaf1995-VANCAC-4}. Games of this kind have played an important role in our understanding of language \cite{sky10}.}

The crucial idea is that signalling turns a belief into public belief. Sometimes, however, beliefs fail to be public. This may be for a number of reasons. In more accidental cases, people are distracted, uncurious, or unintelligent. The impasse here is ``solved'' by a sleight of idealization. Let's assume that Shiv and Logan are Bayesian agents who do not suffer from such accidental shortcomings of rationality. Their credences are mathematically coherent, and they update by Bayes' rule. Still, there are complex cases of failure of public belief known to the literature.

In the following scenario, Shiv and Logan are planning for dinner. However, they are try to coordinate on something vague. The model of vagueness below is discussed in \cite{DeJaegher2003-KRIAGR}, and inspired by the well-known case of two generals' failing to coordinate an attack \cite{Fagin1995-FAGRAK-2,moseshalpern1986}.
\begin{quote}
    \textbf{\textit{Vagueness}}. Shiv and Logan just moved to a new town. They have been told about a great restaurant. If the restaurant is close, they would like to go there for dinner, but since they are quite tired after a day of moving and unpacking, they prefer to eat in if the restaurant is far. They can either go out or stay in, but if either goes while the other doesn't, both will eat alone and be miserable.
\end{quote}
In Vagueness, Shiv and Logan are facing a Lewisian coordination problem. Sometimes, however, a restaurant is neither definitely close, nor definitely far. There is no sharp boundary between close and far, and, in the ``borderline area'', each may think that the restaurant is close while the other thinks that it's far. They will coordinate only if a particular belief is public, but in borderline cases, neither believes that they believe the same thing. Vagueness undermines the possibility of sharing a belief in public.

It would be natural for Shiv and Logan to be uncertain about what to do, in their situation. Vagueness has often been linked to uncertainty \cite{Edgington1992-EDGVUA,MacFarlane2016-MACIVA-4}. A common way to describe uncertainty is in terms of degrees of confidence. Let's say that an agent $i$ \textit{thinks} that $q$ just in case $i$ has some positive degree of confidence that $q$ is the case. Then $i$ thinks that $q$ just in case $p(q_i)>1-p(q_i)$, i.e.\ $i$ expects that $q$ is more likely than not. In the borderline area, Shiv may think that the restaurant is close, although her confidence remains low: indeed, below a relevant threshold. Above the threshold, Shiv thinks that the restaurant is definitely close, or, as I shall say, she \textit{believes} that it is close.

What is a confidence threshold? In ordinary life, many factors contribute to an agent's confidence level. In the context of the game, a qualitative characterization helps: an agent believes that $q$ just in case she thinks that $q$, and thinks that others think that $q$ as well. That is, one believes that $q$ just in case one is confident enough that $q$ is the case to think that others think that $q$ as well. And so someone who thinks that the restaurant is close is uncertain so long as she has a reasonable expectation that others are not of the same opinion.\footnote{Of course, this is not an analysis of \textit{think} and \textit{believe}, but a stipulation for describing a probability distribution over propositions. The stipulation plausibly fits with at least some informal uses of \textit{think} and \textit{believe}. The terminology, however, may sound misleading. Does belief imply certainty? Not if \textit{certainty} means `lack of Cartesian doubt', but it might if it means `having high-ish confidence about what others think'. I mean the latter. Likewise, there is a sense in which one could be confident that something is the case while believing that others disagree: but this just shows that there are other characterizations of what confidence thresholds are, besides what I offered. That's fine. It's worth keeping in mind that Shiv and Logan are supposedly rational epistemic peers: if they think that $q$ expecting that someone else does not think that $q$, who is equally rational and in the same epistemic position, then they should be less confident about their judgement. I assume that they are. That's the relevant sense of \textit{confidence}.}

By assuming that confidence thresholds and shared attitudes line up, the failures of coordination in Vagueness are failures to share a belief about what others think. In other words, coordination fails because a belief isn't public. To make this point precise, let there be at least three discrete states $w_1, w_2, w_3$ in the agents' environment. In $w_1$ the restaurant is close ($q$), in $w_3$ it is far ($\bar{q}$), and in $w_2$ it is neither close nor far. As far as the agents know, prior to the interaction, any of these worlds might be theirs.

\begin{figure}[htb!]
\centering
\footnotesize
\begin{tikzpicture}[scale=0.8]
\tikzstyle{level 1}=[level distance=12mm]
\node (0) at (-0.25,0.5) {$\phi$};
\node (1) at (-0.25,-1) {$\neg\phi$};

\node (6) at (-4,-0.75) {};
\node (16) at (4.5,0.75) {};

\draw (2,1) circle [radius=0.4cm] {};

\node (2) at (-2,1) {$w_1$};
\node (3) at (-2,-1) {$w_3$};
\node (15) at (-3,0) {$S$};
\node (4) at (-2,0) {$w_2$};

\node (12) at (2,0) {$w_2$};
\node (13) at (2,-1) {$w_3$};
\node (14) at (2,1) {$w_1$};
\node (16) at (3,0) {$L$};

\draw (-2,-1) circle [radius=0.4cm] {};

\draw[rounded corners] (-2.4,-0.4) -- (-1.6,-0.4) -- (-1.6,1.4) -- (-2.4,1.4) -- cycle;

\draw[rounded corners] (2.4,-1.4) -- (2.4,0.4) -- (1.6,0.4) -- (1.6,-1.4) -- cycle;

\draw[->] (-1.6,0.5) -- (0);
\draw[->] (3) -- (1);
\draw[->] (1) -- (1.6,-1);
\draw[->] (0) -- (14) node[pos=0.5,above]{$1-\gamma$};
\draw[->] (0) -- (1.6,-0.1) node[pos=0.5,below]{$\gamma$};

\end{tikzpicture}
\caption{Signalling game under uncertainty}
\label{gamesignals}
\end{figure}
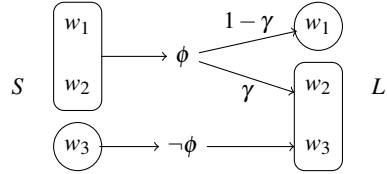

Crucially, the agents' doxastic states are not aligned in the borderline area. For concreteness, let's assume that Shiv does not distinguish $w_1$ and $w_2$, in which the restaurant looks close to her, and Logan does not distinguish $w_2$ and $w_3$, in which the restaurant looks far to him. (The converse possibility is analogous, and omitted.) Thus, the agents partition the logical space differently. A partition is a set of jointly exhaustive and mutually exclusive subsets of the logical space $W$. Each agent $i$ has their own partition $\Pi_i$, which represents how they distinguish possibilities. Let $\pi_i(w)\in\Pi_i$ be the cell of $\Pi_i$ to which $w$ belongs. An agent $i$ `fails to distinguish' $w$ from any $w'$ that belongs to $\pi_i(w)$. I will say that $i$ thinks that $q$ at $w$ just in case $\pi_i(w)\subseteq q$. Figure \ref{gamesignals} represents the agents' doxastic states with respect to the three possibilities that are salient to them at the beginning of the interaction. Furthermore, I assume that Figure \ref{gamesignals} is how the agents understand their own beliefs as well as those of the others.

Partitions in Figure \ref{gamesignals} are generated as follows. Let's consider a Sorites series \texttt{S} of states $t_1,\ldots,t_n$. The restaurant is definitely close in $t_1$ (it is downstairs), it is definitely far in $t_n$ (it is two time zones away), and the remaining states form a linearly ordered progression from near to far. Shiv and Logan go through a `forced march' \cite{Horgan1994-HORRVA}. For all states in \texttt{S}, they judge whether the restaurant is close, $q$, or far, $\bar{q}$. The judgment has to be made even if the agents hesitate. Both Shiv and Logan think that $q$ in $t_1$ since the restaurant is definitely close. For some $x$ between 1 and $n$, Shiv will presumably flip and think that $\bar{q}$ in $t_x$. For some $y$ between 1 and $n$, Logan will flip too. There comes a point during the forced march at which they both judge `It's far', not necessarily the same point. Thus, both Shiv and Logan think that $q$ in all $t<\textnormal{min}(t_x,t_y)$, and both think that $\bar{q}$ in all $t>\textnormal{max}(t_x,t_y)$. Therefore, we may pool \texttt{S} into three ``uber'' states $w_1,w_2,w_3$ without loss of generality, and continue working with uber states (or worlds).
\begin{align*}
    w_1 &=\{t\in\texttt{S}:t<\textnormal{min}(t_x,t_y)\}\\
    w_2 &=\{t\in\texttt{S}:\textnormal{min}(t_x,t_y)\leq t\leq\textnormal{max}(t_x,t_y)\}\\
    w_3 &=\{t\in\texttt{S}:\textnormal{max}(t_x,t_y)<t\}
\end{align*}
The picture could be complicated by adding more players, each drawing the line between $q$ and $\bar{q}$ at a different point. It would still be possible to pool all states in \texttt{S} into those in which $q$ is true for every player, those in which $\bar{q}$ is true for every player, and the rest. Thus, there is a supervaluational description of the Vagueness game environment, in which players take the place of `precisifications' \cite{DeJaegher2003-KRIAGR}.

Finally, let's suppose that Shiv and Logan have two signals, $\phi$ and $\neg\phi$: `The restaurant is close' and `The restaurant is far'. For the semantics, let $\phi$ be true at $w_1$ and false at $w_3$, so $\neg\phi$ is false at $w_1$ and true at $w_3$. I prefer to remain neutral on further details of the semantics of vague terms, such as `close' and `far'. In order to be more specific, one could say that $\phi$ and $\neg\phi$ are truth-valueless at $w_2$. Alternatively, one could keep classical logic, following \cite{Williamson1994-WILV}. My discussion does not depend on the logic of vague terms.

Sending signals reveals to the signal receiver what the signal sender thinks, and this is how coordination is ordinarily reached, if it is. For if they both think the same then, through signalling, they both come to believe that they both think the same. But uncertainty can undermine coordination. Suppose that the agents know that their interaction is as depicted in Figure \ref{gamesignals}. Suppose first that Shiv thinks that the restaurant is far. Then she sends $\neg\phi$. Upon receiving $\neg\phi$, Logan thinks that Shiv thinks that the restaurant is far, for Shiv has no reason to deceive Logan. Therefore, a signal $\neg\phi$ is how Logan can distinguish a possibility in which Logan thinks that the restaurant is far and Shiv does too, from one in which he thinks that the restaurant is far but Shiv doesn't. But Logan thinks that the restaurant is far in any circumstance in which Shiv does, hence if $\neg\phi$ is sent, Shiv and Logan believe that it's far. Moreover, since they can reason to this point, they believe that they believe that it's far, and so on. Therefore, if $\neg\phi$ is sent, the belief that the restaurant is far becomes public, and they coordinate on staying in.

If instead Shiv thinks that the restaurant is close, she sends $\phi$, but public belief might fail. For Shiv fails to distinguish a possibility in which both think that the restaurant is close from another possibility, $w_2$, in which she thinks that the restaurant is close but Logan doesn't. These cases can be interpreted as those in which Shiv wrongly guesses what Logan thinks. For Shiv's signal $\phi$ (`The restaurant is close!') would come across to Logan as an error of judgement: Logan would think that the restaurant is far, and realize by Shiv's signal that she thinks that it's close. Since Shiv is aware of this, if she thinks that the restaurant is close, she may not think that Logan thinks it is.

Coordination equilibria still exist, although within the limits given by the agents' uncertainty. Let $\gamma$ be the agents' shared prior concerning the chance that they give different judgments as to whether the restaurant is close or far, and let $\delta$ be a real number between 0 and 1. The relevant possibilities can then be evaluated as follows.
$$p(w_1)=\delta(1-\gamma) ~ \quad ~ p(w_2)=\gamma ~\quad ~ p(w_3)=(1-\delta)(1-\gamma)$$
That is, with chance $\gamma$ the agents guess that they don't think what the other thinks, i.e., they are in $w_2$, and they assign complementary probabilities to the rest of the cases by splitting them over $\delta$ and $1-\delta$.\footnote{Probabilities are only assigned to sets of possible worlds, in order to uniformly represent an agent's credal state. Therefore, $p(w)$ is strictly speaking a function from the singleton $\{w\}$ to a real number, not a function of a world.} Under the plausible assumption that an agent $i$ chooses $a$ only if $i$ thinks $q$, we can calculate the expected utility of $a$ for $i$. Let $u_i$ be $i$'s utility function, and $j$ be $i$'s opponent. 
$$eu_i(a)=p(q_i~\&~q_j)\cdot u_i(a,a)+p(q_i~\&~\bar{q}_j)\cdot u_i(a,b)$$
By Table \ref{coord}, $u_i(a,b)=0$ for both agents, hence by Bayes' rule,
$$eu_i(a)=p(q_i)\cdot p(q_j|q_i)\cdot u_i(a,a)$$
Consider $S$ first. The probability $p(q_S)$ that $S$ thinks that $q$ is $\delta(1-\gamma)+\gamma$, and the conditional probability $p(q_L|q_S)$ that $L$ thinks that $q$ while $S$ thinks that $q$ is just the proportion of cases in which $L$ thinks $q$ out of those in which $L$ does: $\frac{\delta(1-\gamma)}{\delta(1-\gamma)+\gamma}$. Hence $eu_S(a)=\delta(1-\gamma)$.

Consider $L$. The probability that $L$ thinks that $q$ is $p(w_1)=\delta(1-\gamma)$, and the probability that $S$ thinks that $q$ given that $L$ thinks that $q$ is just $1$, for all cases in which $S$ thinks that $q$ are cases in which $L$ does too. It follows that $eu_L(a)=\delta(1-\gamma)=eu_S(a)$. Parallel reasoning shows that $eu_S(b)=eu_L(b)=(1-\delta)(1-\gamma)$. Consequently, the coordination equilibrium $(a,a)$ obtains only if $eu_S(a)>eu_S(b)$ and $eu_L(a)>eu_L(b)$, hence only if $\delta>1-\delta$. Similarly, the $(b,b)$ equilibrium obtains only if $1-\delta>\delta$.

I have assumed in the preceding paragraph that $\gamma\not=1$. This seems reasonable, since $\gamma$ represents the chance that an agent doesn't think as the other does. In other words, so long as doxastic misalignment isn't inevitable, coordination equilibria exist under the conditions just derived. While reasonable, this conclusion is not very strong. For even if the agents are rational, and know by the proof above that coordination equilibria exist, it doesn't follow that they will coordinate. The uncertainty may still be too impressive for them to take action.

An upper bound on $\gamma$ would help. Earlier I assumed that a necessary condition for $i$ to choose $a$ is that $i$ thinks that $q$. It seems plausible to say that a sufficient condition for $i$ to choose $a$ is that both think that $q$, i.e., that both consider $q$ more likely than not. That is, $a$ is a best response for both if $p(q_i~\&~q_j)>1/2$. Then $a$ is played if $\delta(1-\gamma)>1/2$, i.e., if
\begin{align*}
    \tag{Confidence Threshold for $a$} & \gamma<1-\frac{1}{2\delta}
\end{align*}
By similar reasoning, $b$ is played if
\begin{align*}
    \tag{Confidence Threshold for $b$} & \gamma<1-\frac{1}{2(1-\delta)}
\end{align*}
These inequalities are the confidence thresholds for coordination on $(a,a)$ and $(b,b)$ respectively. They are derived assuming that $\delta$ is neither 0 nor 1, but the generality of the conclusion is not lessened. Such values would trivialize the interaction, for $\delta=0$ would mean that the agents do not consider world $w_1$ a genuine possibility, and on the other hand $\delta=1$ would mean that $w_3$ is not a genuine possibility, given how $p(w_1)$ and $p(w_3)$ were defined above.

The value of $1/2$ as tipping point for action has been chosen somewhat arbitrarily, but the conclusion is representative of a general point. Two conditions characterize the existence of a coordination equilibrium in conditions of uncertainty. For the $(a,a)$ outcome, it must be that $\delta>1-\delta$ and  $\gamma<1-\frac{1}{2\delta}$; for $(b,b)$, that $1-\delta>\delta$ and $\gamma<1-\frac{1}{2(1-\delta)}$. Uncertainty undermines the agents' confidence that something is the case, so that a belief fails to be public. Nevertheless, if the chance $\gamma$ that their thinking differently is not too high, coordination may still obtain. The inequalities CT$a$ and CT$b$ specify what ``not too high'' means. 

\section{Coordination in Times of Uncertainty}\label{assertion}
Confronted with a failure to coordinate beliefs, rational agents could change their mind, of course. However, revising judgements doesn't eliminate vagueness: Shiv and Logan would simply go one step further in the forced march. This may be as good as it gets, if the agents' common language is indeed limited to $\phi$ and $\neg\phi$. Shiv and Logan will then have to learn to live with the occasional failures of coordination. On the other hand, if the agents' language includes epistemic vocabulary, then they could make their uncertainty manifest, and this potentially matters for their attempt to coordinate. To characterize this idea, I will begin with a standard relational semantic for \textit{might}.

The idea that more is communicated in conversation than the semantic content of what the interlocutors say goes back to H. P. Grice \cite{Grice1975-GRILA-2}. Gricean reasoning has a strategic nature, and an appreciation of this point has led to a more systematic game-theoretic understanding of it \cite{clark1996,franke11, fjr12,djvr14,benz2018,Parikh2019-PARCAC-14}. Furthermore, recent work has emphasized the connection between Gricean reasoning and more general Bayesian models of inference under uncertainty that have wide applications in the study of human cognition \cite{Goodman2013-GOOKAI,Goodman2016-GOOPLI-2,Lassiter2017-LASAVI}. The result is a framework for probabilistic inference and back-and-forth reasoning whose outline I will follow in the next sections. 

Suppose that, besides the two signals $\phi$ and $\neg\phi$, the agents' language includes epistemic vocabulary. They can utter sentences such as `The restaurant might be close' and `The restaurant might be far', namely $\Diamond\phi$ and $\Diamond\neg\phi$, respectively. The general idea is that a sentence $\Diamond\phi$ is true just in case $\phi$ is compatible the information some agent has. Roughly, such information is an agent's evidence, or doxastic mental state. For simplicity we may take the relevant agents to be the participants in the game, though of course this would be implausible for the purposes of natural language semantics. If so, then $\Diamond\phi$ is true at a world $w$ in the game model of Figure \ref{gamesignals} just in case there is some agent $i$ who thinks that $\phi$ in $w$. By this light, in $w_2$ it is true to say, `It might be that $\phi$ and it might be that $\neg\phi$'. This seems the right thing to say when one is uncertain, as Shiv and Logan are in $w_2$.

More formally, we define a semantic model over the game of Figure \ref{gamesignals}. The model $(I,W,\Pi,\llbracket\cdot\rrbracket^{c,g})$ includes a set $I$ of players, a set $W$ of worlds, an interpretation function $\llbracket\cdot\rrbracket^{c,g}$ relative to the context and a variable assignment (superscripts henceforth omitted), and a set of partitions $\Pi=\{\Pi_i:i\in I\}$ of $W$, one partition for each player. A rough but standard Kratzerian semantics for $\Diamond$ can be given in terms of $\Pi$ \cite{kra12,vonFintel2011-VONMMR}.
\begin{align*}
   \llbracket \Diamond\phi\rrbracket &=\lambda w.\exists i\in I.\exists \pi_i\in\Pi_i.\exists w'\in \pi_i(w): w'\in \llbracket\phi\rrbracket
\end{align*}
Intuitively, $\Diamond\phi$ is true at $w$ iff there is a $w'$ accessible from $w$ such that $\phi$ is true at $w'$. A world is accessible form another just in case they belong to the same cell of some agent's partition. Thus, epistemically accessible worlds are those that some agent finds indistinguishable on the basis of their doxastic perspective prior to communication. It is straightforward to check that accessibility, thus defined in terms of doxastic partitions, is reflexive, symmetric, and not transitive. 

In order to calculate the pragmatic effects of manifesting uncertainty by asserting that $\Diamond\phi$, let's refer to the set $\{w_1,w_2,w_3\}$ as Shiv and Logan's \textit{common ground} at time 0, $cg(0)$: the worlds the interlocutors jointly consider to be possible, at the beginning of their interaction. Following Stalnaker  \cite{sta78,sta99,sta02}, conversation is a cooperative enterprise whereby interlocutors narrow down the common ground. The task for the listener is to figure out which world is actual, given what the speaker said. A simple hypothesis is that worlds in the common ground, at any time, have equal chances of being actual. On the basis of this hypothesis, base-rate probabilities may be easily calculated for any time $t$.
$$\textnormal{For all times $t$ and for all $w$ in }cg(t): p(w)=\frac{1}{|cg(t)|}$$
Therefore, once the agents narrow down the common ground to $\{w\}$, the probability that $w$ is the actual world is 1. In $cg(0)$, we have $p(w_1)=p(w_2)=p(w_3)=1/3$.

Suppose for illustration that Shiv thinks the restaurant is close, but can't tell if Logan thinks so as well. For all she knows (we could say, semantically ascending), the actual world is $w_1$ or $w_2$: after all, she cannot distinguish these two possibilities. In both $w_1$ and $w_2$, $\Diamond\phi$ is true, as for both worlds there is an agent, namely Shiv, who thinks that the restaurant is close at those worlds. Therefore (semantically descending), Shiv believes that $\Diamond\phi$ is true. Thus, she asserts so. An assertion is a proposal to update the common ground, by eliminating possibilities that are incompatible with the semantic content of the assertion \cite{sta78}. By the standard semantics I assumed above, and the Stalnakerian dynamics of assertion, an assertion of $\Diamond\phi$ rules out $w_3$, which is incompatible with the truth of the assertion. After Shiv's assertion that $\Diamond\phi$, the possibility that the restaurant is definitely far is no longer relevant for either Shiv or Logan.

Logan may now reason that if Shiv had meant to suggest that $w_1$ is the actual world, she would have sent $\phi$ (`The restaurant is close!') right away, for $w_1$ is the world in which both think that the restaurant is close. But she didn't send $\phi$: she wasn't confident enough for that. So, she doesn't think that the restaurant is definitely close. Since $w_2$ is the only other possibility left, $w_2$ must be the actual world according to the speaker. Thus the agents become aware of a distinction between confidence levels by using epistemic language.

This reasoning can be formalized in a Bayesian framework. At time $1$, after the update, the listener $L$ has equal priors for the worlds in $cg(1)$, i.e.\ $p(w_1)=p(w_2)=1/2$. Moreover, $L$ expects $S$ to be truthful. Since a truthful speaker could send only $\phi$ and $\Diamond\phi$ in $cg(1)$, $L$ holds even priors for the events that these signals are sent, i.e.\ $p(\phi)=p(\Diamond\phi)=1/2$. Finally, $L$ expects $S$ to send $\phi$ in $w_1$, not in $w_2$. For an assertion that $\phi$ reveals the speaker's belief that the restaurant is close, but the speaker believes that the restaurant is close only in $w_1$. Therefore, $L$'s conditional probability for the event that $\phi$ is sent given that $w_1$ is the actual world is at least nearly 1. Conversely for $\Diamond\phi$.
\begin{align*}
 p(\phi|w_1)\approx 1&\qquad p(\Diamond\phi|w_1)\approx 0\\
 p(\phi|w_2)\approx 0&\qquad p(\Diamond\phi|w_2)\approx 1
\end{align*}
The last step is for $L$ to update by Bayes rule. The posterior probability that a world is actual is calculated by the listener by conditionalizing on the evidence, namely the observation that $\Diamond\phi$ was sent.
$$p'(w_2)=\frac{p(\Diamond\phi|w_2)\cdot p(w_2)}{p(\Diamond\phi)}\approx\frac{1\cdot 1/2}{1/2}\approx 1$$
From the observation that `It might be raining' was uttered, with the semantics it has, and given what else could have been uttered, the listener draws a conclusion about the speaker's confidence level: in the actual world the restaurant is neither definitely close nor definitely far, and in particular the speaker thinks that it's close but she is not confident. The listener's inference is a defeasible one, and not a semantic entailment. Like ordinary pragmatic reasoning, its conclusion is not packaged in the semantic content of the sentence that was uttered by the speaker. Thus, the sentence `It might be raining' is not about the speaker's credal state (or anybody else's, for that matter). Yet it supports a Bayesian inference to a conclusion about the speaker's credences.

\section{Strategic Hedging}

How do rational agents react to someone's assertion that the restaurant might be close? Earlier I assumed that an agent goes out if they think that the restaurant is close, and stays in if they think that it's far. In $w_2$ the restaurant is neither close nor far, and Shiv thinks that it's close while Logan thinks that it's far. Consequently, they don't coordinate. The assumption is rather crude, however, for we might want to say that uncertainty comes with indecision \cite{MacFarlane2016-MACIVA-4}.

If Shiv says `The restaurant might be close', she signals her uncertainty to Logan, who infers it as a good Bayesian. Moreover, Shiv might expect this inference to be of some consequence. Logan would have to take Shiv's uncertainty into account. At the very least, Shiv might expect that Logan hesitates before taking action, once `Might $\phi$' is asserted. I will assume that she does. I will now show that, as a consequence, it's reasonable for Logan to go out, when he is told that the restaurant might be close, even if he thinks that it's far. That is, the chances of coordination improve despite failure of public belief.

Shiv's expectation about Logan's reaction to her utterance kick-starts an expectation-building process. For she will have higher-order expectations about what her reaction will be to what she expects Logan's reaction to her utterance is, and so on. The result is essentially an instance of iterated reasoning between speaker and listener. Taking notice of each other, the interlocutors adjust their propensity to act.\footnote{There are two slightly different frameworks one could use to reconstruct this process: \textit{iterated best response} models of pragmatic reasoning \cite{franke11}, and \textit{rational speech act} models \cite{frank2017}. The discussion in this section is inspired mainly by the latter, but could be carried out in the former setting with some adjustments.}

Shiv and Logan's reasoning about each others' actions takes place under the assumption that their mental states are incompatible. In other words, we are in $w_2$, and the agents correctly inferred this by Bayesian reasoning as above. Since the agents don't change their mind concerning the restaurant's location as they go through the expectation-building process below, probabilities are normalized at each step. Thus, their mental states remain incompatible throughout, insofar as beliefs can be surmised by dispositions to act. Nevertheless, we will see that Shiv and Logan's expected utilities increase. I indicate with $p_i(x)$ the probability that agent $i$ performs action $x$, and break down the reasoning in several steps.\footnote{More precisely, $p_i(x)$ is short for $p_i(x|q_S~\&~\bar{q}_L)$: the conditional probability that $i$ does $x$ given that $S$ thinks that $q$ and $L$ thinks that $\bar{q}$. We are holding fixed that we are in $w_2$, in which the condition $q_S~\&~\bar{q}_L$ holds.}
\begin{itemize}
    \item[] \textbf{Step 0}: prior to the use of epistemic modals. $S$ thinks that $q$, so she chooses $a$. This fixes the speaker's prior, which is $p_S$ at step 0. At the same time, $L$ thinks that $\bar{q}$, so he chooses $b$. Coordination at this stage inevitably fails.
$$p^0_S(a)=1\qquad p^0_S(b)=0$$
$$p^0_L(a)=0\qquad p^0_L(b)=1$$
    \item[] \textbf{Step 1}: using epistemic modals. $S$ signals $\Diamond\phi$ and expects that $L$ hesitates. $L$'s expected hesitation is a matter of randomly choosing $a$ or $b$. This fixes the listener's prior, which is $p_L$ at step 1.
       $$p^1_L(a)=p^1_L(b)=0.5$$
    \item[] \textbf{Step 2}: expectation-building. $S$ reflects on her action in response to the listener's prior. Each next step from now on is obtained by normalizing an agent's prior with the other player's.
    $$p^1_S(a)=\frac{p^0_S(a)}{\Sigma_{i\in I}p_i(a)}=\frac{p^0_S(a)}{p^0_S(a)+p^1_L(a)}=\frac{1}{1+0.5}\approx0.666$$
    
\item[] \textbf{Step 3}: as in the previous step.
$$p^2_L(a)=\frac{p^1_L(a)}{p^1_L(a)+p^2_S(a)}=\frac{0.5}{0.5+0.666}\approx0.428$$
\end{itemize}
By proceeding in this way, the probability that $S$ chooses $a$ in $w_2$ tends approximately to $0.6$, and the probability that $L$ does so tends approximately to $0.4$. Conversely for $b$. The step-wise process can be summarized by a system of equations. This is an inductive definition of a function $f_a(n)$ that maps a number $n$ that counts the steps, to the probability that an agent takes action $a$ at step $n$. The probability of doing $a$ for the speaker is given by $f_a(2n)$, i.e., for steps indexed by even numbers, whereas the probability of doing $a$ for the listener is given by $f_a(2n+1)$. A similar series can be defined for $b$. 
\begin{align*}
    f_a(0) &=1\\
f_a(1) &=1/2\\
f_a(n) &=\frac{f_a(n-2)}{f_a(n-1)+f_a(n-2)}
\end{align*}
$f_a$ defines a divergent sequence of probabilities, oscillating between approximately $0.4$ and approximately $0.6$. This can be observed by simple calculation. Analytic proof is quite involved, and left out of the paper.

While they go through the (first few steps of the) stepwise process, the agents' dispositions to act are incompatible throughout, as an effect of normalizing probabilities. However, the margin by which such incompatibility causes failures of coordination is reduced with each step. So, they still share no public belief, but their expected utility is higher. Recall that the value of an agent's expected utility for action $a$, as calculated above, was:
$$eu_i(a)=p(q_i~\&~q_j)\cdot u_i(a,a)=\delta(1-\gamma)$$
However, this equation assumes that one gets a payoff for $a$ just in case both think that $q$. But one's payoff for $a$ increases, via $f_a$, also in proportion to the probability that $S$ thinks that $q$, $L$ thinks that $\bar{q}$, but both do $a$. So we revise the notion of expected utility, indexing it to the number of steps.
\begin{align*}
    eu^n_i(a) &= eu_i(a)+p(q_S~\&~\bar{q}_L)\cdot p^n_S(a)\cdot p^n_L(a)\cdot u_i(a,a)\\
    &= \delta(1-\gamma) + \gamma\cdot p^n_S(a)\cdot p^n_L(a)\cdot u_i(a,a)
\end{align*}
At Step 0, the listener doesn't think that $q$, thus $p_L^0(a)=0$. Therefore, the overall expected utility of $a$ at 0 is simply $eu_i(a)$, as above. Assuming instead that we are at Step 3:
\begin{align*}
    eu^3_i(a) &=\delta(1-\gamma)+\gamma\cdot 0.666\cdot 0.428=\delta(1-\gamma)+\gamma\cdot 0.285
\end{align*}
More generally, for all actions $x$, and for all $n\geq 0$, the agents' expected utility monotonically increases with the sequence of steps. 
$$eu_i^0(x)\leq eu_i^n(x)$$
This argument is fairly abstract, but it's a mathematical reconstrution of a plausible conclusion. A reasoning process can be defined on the basis of the agents' expectations, in reaction to the uncertainty manifested by an assertion of `Might $\phi$'. The base step of the induction is the intuitively plausible idea that the speaker, having signalled her uncertainty, expects the listener to hesitate before acting. Based on this, the speaker reflects on how to react to the listener's hesitation, on how the listener would react to her reaction, and so on. The agents need not have perfect powers of reasoning. They need not follow the induction to infinity. It suffices that one or two steps are taken, and already the use of $\Diamond\phi$ leads to higher expected utility.

If the restaurant is neither close nor far, going out is reasonable not only for Shiv (who thinks with little confidence that the restaurant is close), but also for Logan (who thinks with little confidence that it's far), in response to the speaker's assertion that it might be close. This choice is \textit{reasonable} in the very concrete sense of expected utility maximization. Thus, by hedging one's assertion in conditions of uncertainty via the use of epistemic possibility modals, the chances of coordination improve although public belief fails. 

\section{Conclusion}

In this paper, I presented a ``proof of concept'' for the use of epistemic modal expressions in signalling games in which uncertainty (about what another player thinks) undermines coordination. Vagueness may trigger uncertainty of this kind, since it undermines the belief that others think in the same way as we do. However, by using `Might $\phi$', we hedge our assertions and make uncertainty manifest. This can be seen by a straightforward application of Bayes' rule, on the basis of a standard semantics for \textit{might} and the Stalnakerian pragmatics of assertion as informative update. In turn, a manifestation of uncertainty may lead interlocutors to accommodate their actions with what they expect the others' actions will be, even though their doxastic mental states remain incompatible throughout. Coordination under uncertainty is facilitated by the strategic assertion of `Might $\phi$'.

By necessity, the view I presented applies only to particular contexts, formalized as particular kinds of games. By no means I suggest that the interaction I described is the only effect epistemic possibility modals have in an interactive setting. The semantics for epistemic modality I adopted is somewhat rough but standard, and could be fine-tuned for the purposes of natural language semantics. The rational speech act model I adopted is an abstract formalization of the computational import of epistemic signalling, but could be understood as an element of a cognitively plausible picture of bounded rationality in interaction. 

\bibliographystyle{eptcs}
\bibliography{bibliography}
\end{document}